\documentclass[preprintnumbers,amsmath,amssymb,superscriptaddress]{revtex4}
\usepackage{bm}
\usepackage{graphicx}
\usepackage{dcolumn}
\usepackage{xcolor}
\usepackage{hyperref}
\usepackage{diagbox}
\usepackage{multirow}
\usepackage{diagbox}
\usepackage{mathrsfs}
\usepackage{pifont}
\begin{document}

\title{Detecting $k$-nonstretchability via a class of informationally complete symmetric measurements}

\author{Yan Hong}
\affiliation{School of Mathematics and Science, Hebei GEO University, Shijiazhuang 052161, China}
\affiliation{Intelligent Sensor Network Engineering Research Center of Hebei Province, Hebei GEO University, Shijiazhuang 052161, China}

\author{Mengjia Zhang}
\affiliation{School of Mathematics and Science, Hebei GEO University, Shijiazhuang 052161, China}

\author{Limin Gao}
\email{gaoliminabc@163.com}
\thanks{Lead contact.}
\affiliation{School of Mathematics and Science, Hebei GEO University, Shijiazhuang 052161, China}
\affiliation{Intelligent Sensor Network Engineering Research Center of Hebei Province, Hebei GEO University, Shijiazhuang 052161, China}

\author{Huaqi Zhou}
\email{zhouhuaqilc@163.com}
\affiliation{School of Mathematics and Science, Hebei GEO University, Shijiazhuang 052161, China}

\author{Limei Zhang}
\affiliation{School of Mathematics and Computer Science, Hengshui University, Hengshui 053000, China}

\begin{abstract}
Characterizing multipartite entanglement is a fundamental problem in quantum information theory. The concept of $k$-stretchability provides a framework for characterizing the structure of multipartite entanglement. We investigate $k$-nonstretchability using informationally complete $(s,t)$-positive operator-valued measures ($(s,t)$-POVMs) and derive two families of criteria. These criteria identify classes of $k$-nonstretchable states, and we demonstrate their applicability and advantages through explicit examples.
\end{abstract}

\maketitle

\section{Introduction}

Quantum entanglement is one of the most fundamental features of quantum physics \cite{HorodeckiRMP2009,GuhneToth2009}, with important applications in quantum teleportation, quantum computation, and quantum communication \cite{BennettWiesnerPRL1992,BennettBrassardCrepeauJozsaPeresWoottersPRL1993,ShorSIAMReview1999}. The quantification and detection of entanglement have long been central topics in quantum information theory.

For bipartite quantum systems, significant progress has been made, leading to the development of various criteria, including the PPT criterion \cite{Peres1996}, the
reduction criterion \cite{HorodeckiHorodeckiPRA1999}, the realignment criterion \cite{ChenWu2003,Rudolph2003}, as well as criteria based on extension methods  \cite{DohertyParrilo2002PRA}, covariance matrices \cite{GuhneHyllusGittsovichEisert2007,GittsovichGuhneHyllusEisert2008}, quantum Fisher information \cite{LiLuo2013PRA}, among others.

Owing to the complexity of multipartite quantum systems, various approaches have been proposed for classifying multipartite entanglement, among which $k$-separability and $k$-producibility are the most widely used ones \cite{HorodeckiRMP2009,GuhneToth2009}. Recently, a novel classification framework called $k$-stretchability has been introduced to characterize the structure of multipartite entanglement \cite{Szalay2019Quantum}. Although $k$-separability, $k$-producibility, and $k$-stretchability represent distinct frameworks for an $N$-partite quantum system $H_1\otimes H_2\otimes\cdots \otimes H_N$, they are all fundamentally based on different notions of partitions of the subsystem set \cite{Szalay2019Quantum,RenLiSmerziGessnerPRL2021,SzalayToth2025Quantum}.

To further advance the understanding of multipartite entanglement, numerous methods have been developed, leading to substantial progress in entanglement characterization. Several criteria have been proposed, including $k$-nonseparability criteria based on quantum Fisher information \cite{HongLuoSong2015}, nonlinear local operators \cite{GaoYan2014,EPL104.20007,HongGaoYanPLA2021}, Greenberger-Horne-Zeilinger-class fidelity \cite{PlodzienChwedenczukLewensteinMieldzio2024}, $k$-entanglement detection criteria based on Wigner-Yanase skew information \cite{ChenPRA2005}, quantum Fisher information \cite{Hyllus2012,Goth2012,AkbariAzhdargalam2019}, semidefinite programming \cite{WuChen2025}, and a $k$-nonstretchability detection criterion based on quantum Fisher information \cite{RenLiSmerziGessnerPRL2021}. Nevertheless, these criteria are not sufficient to fully characterize the structure of multipartite entanglement, motivating the development of additional tools for constructing more effective entanglement detection methods.

As two types of symmetric measurements, generalized symmetric informationally complete positive operator-valued measures (GSIC-POVMs) and mutually unbiased measurements (MUMs) have been extensively investigated \cite{KalevGour2014,Rastegin2015,GourKalev2014,Rastegin2014,WiesniakPaterekZeilinger2011,DurtBengtssonZyczkowski2010,Appleby2007,
ChenMaFeiPRA2014,ShenLiDuanPRA2015,LiuGaoYan2018,ShangAsadianPRA2018,SpenglerHuberPRA2012,LaiLuo2022}.
More recently, a broader class of informationally complete symmetric measurements, called informationally complete $(s,t)$-POVMs, has been introduced
\cite{SiudzinskaPRA2022}, with GSIC-POVMs and MUMs as special cases. Since their introduction, $(s,t)$-POVMs have been applied to entanglement
characterization. Specifically, Refs. \cite{WangSunFeiWangQIP2024,TavakoliMorelliPRA2024} developed methods for determining the Schmidt number
using $(s,t)$-POVMs, while Refs. \cite{LiYaoFeiFanMaPRA2024,WangFeiPRA2025} employed them to establish lower bounds on concurrence. Furthermore,
$(s,t)$-POVMs have been used to identify entangled states \cite{SiudzinskaPRA2022,SchumacherAlberPRA2023,LaiLuoCommunTheorPhys.2023,TangWuPhysScr2023,QiPangHouQIP2025,SiudzinskaSciRep2022}. These results motivate further investigation of multipartite entanglement via informationally complete $(s,t)$-POVMs.

In this work, we establish two classes of $k$-nonstretchability criteria for multipartite quantum systems based on informationally complete $(s,t)$-POVMs. These criteria can identify classes of $k$-nonstretchable states. In Sec. II, we review the concepts and properties of informationally
complete $(s,t)$-POVMs, the metric-adjusted skew information, and the variance. In Sec. III, we derive two classes of $k$-nonstretchability criteria
via informationally complete $(s,t)$-POVMs. In Sec. IV, we illustrate the effectiveness of these criteria through concrete examples.

\section{Preliminaries and notation}

In this section, we introduce the concepts and properties of $k$-stretchability, informationally complete $(s,t)$-POVMs, the metric-adjusted skew information, and the variance, which form the foundation for deriving the $k$-nonstretchability criteria in the subsequent sections.

\subsection{$k$-stretchable states}

The classification of entanglement in multipartite quantum systems is a highly complex topic. Currently, the dominant approaches classify multipartite entanglement either according to the number of partitions (i.e., $k$-separability) or according to the number of entangled particles (i.e., $k$-partite entanglement) \cite{GuhneToth2009}. However, these approaches cannot fully capture the rich structure of multipartite entanglement. The concept of $k$-stretchability was introduced to address this limitation \cite{Szalay2019Quantum}.

Let $\gamma := \gamma_1|\cdots|\gamma_{|\gamma|}$ be a partition of the set $\{1,2,\cdots,N\}$, where the $\gamma_i$ are nonempty mutually disjoint subsets satisfying
$\bigcup\limits_{1\leq i\leq|\gamma|}\gamma_i=\{1,2,\cdots,N\}$. The partition $\gamma := \gamma_1|\cdots|\gamma_{|\gamma|}$ is called a $k$-stretchable partition if $\max\limits_{1\leq i\leq|\gamma|}n_{\gamma_i}-|\gamma|\leq k$, where $|\gamma|$ denotes the size of the partition and $n_{\gamma_i}$ denotes the number of particles in the subset $\gamma_i$. A $k$-stretchable partition can be represented by a Young diagram with height $h$ and width $w$, where $h$ represents the
size of the partition and $w$ represents the maximum number of particles in a subset, subject to the constraint $w-h\leq k$ \cite{RenLiSmerziGessnerPRL2021}. For example, the partition $\gamma_1|\gamma_2|\gamma_3|\gamma_4$ with $n_{\gamma_1}=2, n_{\gamma_2}=n_{\gamma_3}=n_{\gamma_4}=1$, and $\gamma'_1|\gamma'_2|\gamma'_3|\gamma'_4|\gamma'_5$ with $n_{\gamma'_1}= n_{\gamma'_2}=n_{\gamma'_3}=n_{\gamma'_4}=n_{\gamma'_5}=1$  are both $-2$-stretchable partitions of the set $\{1,2,\cdots,5\}$, as shown in Fig.~1.
\begin{figure}[h]
\centering
\includegraphics[width=0.5\textwidth]{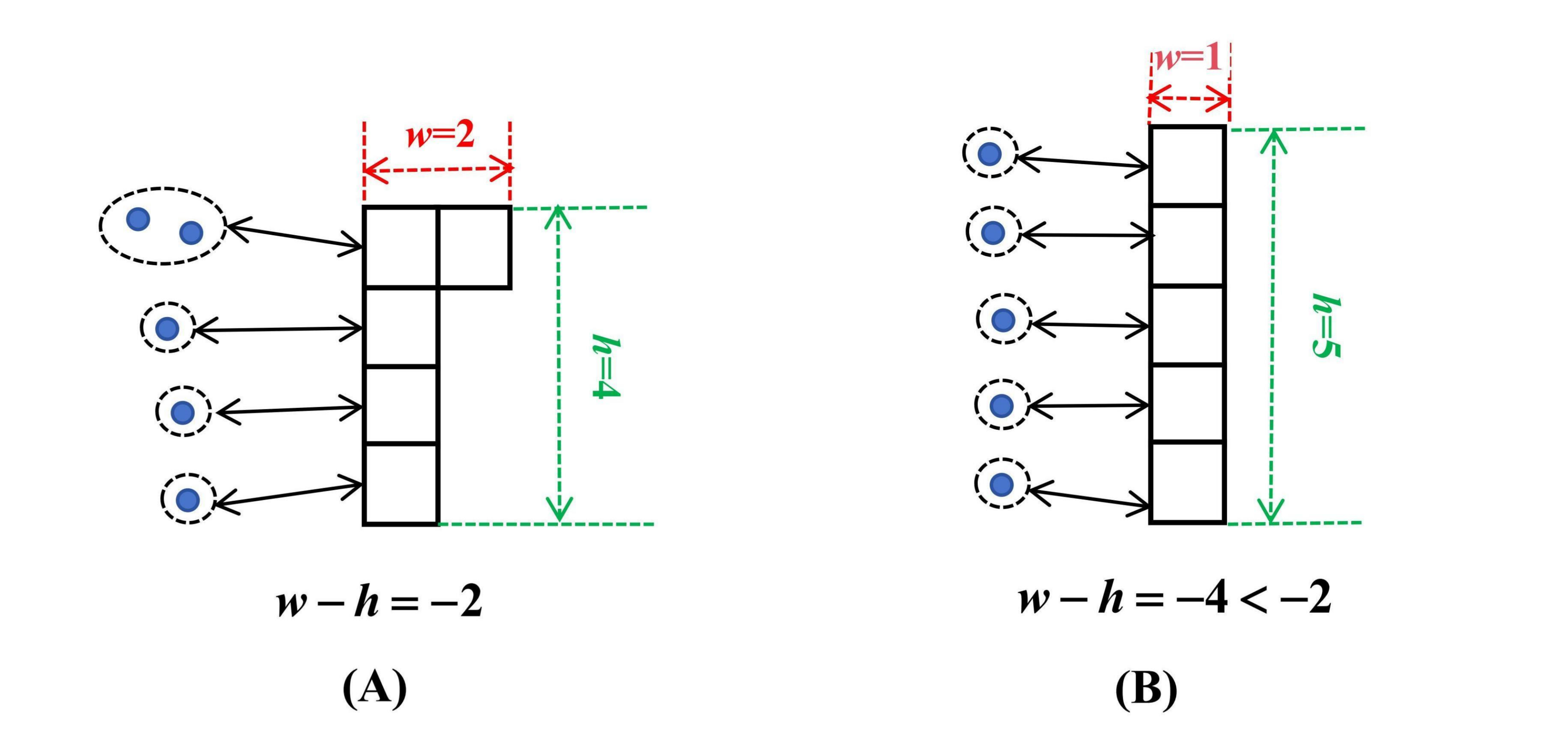}
\caption{These two diagrams represent the Young diagrams corresponding to the $-2$-stretchable partitions of the set $\{1,2,\cdots,5\}$. (A) denotes
the Young diagram corresponding to the partition $\gamma_1|\gamma_2|\gamma_3|\gamma_4$ with $n_{\gamma_1}=2, n_{\gamma_2}=n_{\gamma_3}=n_{\gamma_4}=1$,
and (B) denotes the Young diagram corresponding to $\gamma'_1|\gamma'_2|\gamma'_3|\gamma'_4|\gamma'_5$ with $n_{\gamma'_1}=
n_{\gamma'_2}=n_{\gamma'_3}=n_{\gamma'_4}=n_{\gamma'_5}=1$. }
\end{figure}

In an $N$-partite quantum system $H_1\otimes H_2\otimes\cdots \otimes H_N$, the pure state $|\psi\rangle$ is called $k$-stretchable if it can be written as \cite{Szalay2019Quantum}
\begin{equation}\label{stretchabilitypure}
\begin{array}{rl}
|\psi\rangle=|\psi_{\gamma_1}\rangle|\psi_{\gamma_2}\rangle\cdots|\psi_{\gamma_{|\gamma|}}\rangle,
\end{array}
\end{equation}
where  $\gamma_1|\cdots|\gamma_{|\gamma|}$
is a $k$-stretchable partition of $\{1,2,\cdots,N\}$ and $|\psi_{\gamma_i}\rangle$ is a state on the subsystem $\bigotimes\limits_{j\in \gamma_i}H_j$.
 An $N$-partite mixed state $\rho$ on $H_1\otimes H_2\otimes\cdots \otimes H_N$ is called $k$-stretchable if it can be represented as a convex combination of $k$-stretchable pure states \cite{Szalay2019Quantum},
 \begin{equation}\label{stretchabilitymixed}
\begin{array}{rl}
\rho=\sum\limits_lp_l|\psi^{(l)}\rangle\langle\psi^{(l)}|,
\end{array}
\end{equation}
where each pure state $|\psi^{(l)}\rangle$ is $k$-stretchable, possibly with respect to a different $k$-stretchable partition. If a quantum state $\rho$ is not $k$-stretchable, it is $k$-nonstretchable. In the extreme case, the conventional notion of full separability is recovered: a $(1-N)$-stretchable state is precisely a fully separable state.

$k$-stretchability provides a more refined and comprehensive description of the entanglement structure of the multipartite quantum states considered here  \cite{Szalay2019Quantum,SzalayToth2025Quantum}.
Let $|\psi_1\rangle=\big(\frac{|000\rangle+|111\rangle}{\sqrt{2}}\big)|0\rangle|0\rangle$, $|\psi_2\rangle=\big(\frac{|00\rangle+|11\rangle}{\sqrt{2}}\big)\big(\frac{|00\rangle+|11\rangle}{\sqrt{2}}\big)|0\rangle$,
$|\psi_3\rangle=\big(\frac{|00\rangle+|11\rangle}{\sqrt{2}}\big)|0\rangle|0\rangle|0\rangle$, and
$|\psi_4\rangle=\big(\frac{|00\rangle+|11\rangle}{\sqrt{2}}\big)\big(\frac{|00\rangle+|11\rangle}{\sqrt{2}}\big)|0\rangle$.
Both $|\psi_1\rangle$ and $|\psi_2\rangle$ are 3-separable; thus, in terms of separability, these two quantum states cannot be fully distinguished. Since $|\psi_1\rangle$ is 3-producible whereas $|\psi_2\rangle$ is 2-producible, the two states can be distinguished in terms of producibility. However, producibility only accounts for the width $w$. $|\psi_1\rangle$ is $0$-stretchable, whereas $|\psi_2\rangle$ is $-1$-stretchable. Thus, in terms of stretchability, the two states can also be distinguished, since stretchability accounts for both $w$ and $h$. Both $|\psi_3\rangle$ and $|\psi_4\rangle$ are 2-producible; hence, they cannot be fully distinguished in terms of producibility. However, $|\psi_3\rangle$ is 4-separable whereas $|\psi_4\rangle$ is 3-separable; therefore, the two states can be distinguished in terms of separability. Nevertheless, separability only accounts for the height $h$. Since $|\psi_3\rangle$ is $-2$-stretchable whereas $|\psi_4\rangle$ is $-1$-stretchable, they can also be distinguished in terms of stretchability, which accounts for both $w$ and $h$. However, in some cases, quantum states with different structures may have the same stretchability, leading to a loss of distinguishability.

For example, both $|\psi_5\rangle=\big(\frac{|0000\rangle+|1111\rangle}{\sqrt{2}}\big)|0\rangle|0\rangle$ and
$|\psi_6\rangle=\big(\frac{|000\rangle+|111\rangle}{\sqrt{2}}\big)\big(\frac{|000\rangle+|111\rangle}{\sqrt{2}}\big)$ are $1$-stretchable, although they have different entanglement structures. This illustrates a limitation of using a single stretchability parameter to capture both $w$ and $h$.
Therefore, $k$-stretchability has both advantages and certain limitations in characterizing the structure of multipartite entanglement.

\subsection{Informationally complete $(s,t)$-POVMs}
Quantum measurement is a fundamental mechanism in quantum information processing. Informationally complete $(s,t)$-POVMs form a general framework that encompasses both GSIC-POVMs and MUMs, and provide a versatile and analytically tractable class of symmetric measurements. Their structural generality and mathematical richness make them
particularly valuable for advancing the characterization of multipartite entanglement.

In a $d$-dimensional Hilbert space $H$, a set $\{\mathcal{A}^{(u)}:u=1,2,\cdots,s\}$ of $s$ POVMs
$\mathcal{A}^{(u)}=\{A^{(uv)}:v=1,2,\cdots,t\}$
is called an $(s,t)$-POVM if it satisfies the symmetry conditions \cite{SiudzinskaPRA2022}:
\begin{equation*}
\begin{aligned}
\textrm{Tr}\big(A^{(uv)}\big)&=\dfrac{d}{t}, \\
 \textrm{Tr}\big(\big(A^{(uv)}\big)^2\big)&=\chi, \\
 \textrm{Tr}\big(A^{(uv)}A^{(uv')}\big)&=\dfrac{d-t\chi}{t(t-1)}, \quad v\neq v',\\
 \textrm{Tr}\big(A^{(uv)}A^{(u'v')}\big)&=\dfrac{d}{t^2}, \quad u\neq u',
\end{aligned}
\end{equation*}
where $\chi$ is a free parameter satisfying
$\dfrac{d}{t^2}<\chi\leq\min\Big\{\dfrac{d^2}{t^2},\dfrac{d}{t}\Big\}.$
The set of operators $\{A^{(uv)}:u=1,2,\cdots,s;\ v=1,2,\cdots,t\}$ is an informationally complete $(s,t)$-POVM if and only if $s(t-1)=d^2-1$
\cite{SiudzinskaPRA2022}. In a Hilbert space $H$ with $\dim(H)=d$, there exist at least four classes of informationally complete $(s,t)$-POVMs
\cite{SiudzinskaPRA2022}:
(1) $s=1$ and $t=d^2$; \
(2) $s=d+1$ and $t=d$; \
(3) $s=d^2-1$ and $t=2$; \
(4) $s=d-1$ and $t=d+2$. \

In Ref.~\cite{SiudzinskaPRA2022}, a general construction for informationally complete $(s,t)$-POVM in a Hilbert space $H$ with $\dim(H)=d$ was
presented. Let
$\Big\{\dfrac{\mathbf{1}}{\sqrt{d}}, C^{(uv)}:u=1,2,\cdots, s;v=1,2,\cdots,t-1\Big\}$
be an orthonormal Hermitian operator basis, where each $C^{(uv)}$ is traceless and $\mathbf{1}$ is the identity operator. Define
\begin{equation}\label{symmetricmeasurements1}
\begin{aligned}
B^{(uv)}=\left\{\begin{array}{ll} C^{(u)}-\sqrt{t}(\sqrt{t}+1)C^{(uv)}, &  v=1,2,\cdots,t-1,\\
 (\sqrt{t}+1)C^{(u)},&  v=t,
\end{array}
\right.
\end{aligned}
\end{equation}
where $C^{(u)}=\sum\limits_{v=1}^{t-1}C^{(uv)}$. Then an informationally complete $(s,t)$-POVM is given by
\begin{equation}\label{symmetricmeasurements2}
\begin{aligned}
A^{(uv)}=\dfrac{1}{t}\mathbf{1}+rB^{(uv)},\quad u=1,2,\cdots,s;\ v=1,2,\cdots,t.
\end{aligned}
\end{equation}
The parameter $r$ is chosen such that $A^{(uv)}\geq0$ for all $u$ and $v$. Furthermore, it was shown in Ref.~\cite{SiudzinskaPRA2022} that any informationally complete set of $(s,t)$-POVMs can be obtained by this method.

\emph{Lemma 1.}
Let $\{A^{(uv)}:u=1,2,\cdots,s;\ v=1,2,\cdots,t\}$ be any informationally complete $(s,t)$-POVM.

(i) These operators satisfy the following relation:
\begin{equation}\label{bound0}
\begin{aligned}
\sum\limits_{u=1}^{s}\sum\limits_{v=1}^t\big(A^{(uv)}\big)^2=\Big[\frac{s}{t}+r^2t(\sqrt{t}+1)^2\Big(d-\frac{1}{d}\Big)\Big]\mathbf{1}.
\end{aligned}
\end{equation}

(ii) For any quantum state $\rho$, the following relation holds \cite{SiudzinskaPRA2022}:
\begin{align}
\sum\limits_{u=1}^{s}\sum\limits_{v=1}^t\Big[\textrm{Tr}\big(A^{(uv)}\rho\big)\Big]^2
=&\frac{d(t^2\chi-d)\textrm{Tr}(\rho^2)+d^3-t^2\chi}{dt(t-1)}\label{bound00}\\
\leq&\frac{(d-1)(d^2+t^2\chi)}{dt(t-1)}.\label{bound000}
\end{align}
Equality in (\ref{bound000}) holds when $\rho$ is a pure state.

The proof of Eq.~(\ref{bound0}) is provided in Appendix A. Equation~(\ref{bound00}) and inequality~(\ref{bound000}) are given in Ref.~\cite{SiudzinskaPRA2022}.

\subsection{The metric-adjusted skew information and variance}

Let the spectral decomposition of a quantum state $\rho$ be $\rho=\sum\limits_i\lambda_i|\psi_i\rangle\langle\psi_i|$. Then the metric-adjusted
skew information of an observable $X$ in the state $\rho$ can be written as \cite{SunLiLuo2022}
\begin{equation*}
\begin{aligned}
I_f(\rho,X)=\frac{f(0)}{2}\sum\limits_{i,j}\dfrac{(\lambda_i-\lambda_j)^2}{\lambda_jf(\lambda_i/\lambda_j)}|\langle\psi_i|X|\psi_j\rangle|^2,
\end{aligned}
\end{equation*}
where $f:\mathbb{R}^+\to\mathbb{R}^+$ is an operator monotone function satisfying $f(0)>0$ and $xf(1/x)=f(x)$. Here, $f$ is called operator monotone if $A\leq B$ implies $f(A)\leq f(B)$ for any Hermitian operators $A$ and $B$ \cite{FrankHansen2008,CaiHansen2010}. When $f(x)=\frac{\omega(1-\omega)(x-1)^2}{(x^\omega-1)(x^{1-\omega}-1)}$ or $f(x)=\frac{1+x}{2}$, the metric-adjusted skew information reduces
to the Wigner-Yanase-Dyson skew information and the quantum Fisher information \cite{FrankHansen2008,CaiHansen2010,SunLiLuo2022}, respectively.
The variance of an observable $X$ in state $\rho$ is defined as \cite{HofmannTakeuchi2003}
 \begin{equation*}
\begin{aligned}
V(\rho,X)=\textrm{Tr}\big(\rho X^2\big)-\big[\textrm{Tr}(\rho X)\big]^2.
\end{aligned}
\end{equation*}
Note that \cite{CaiHansen2010}
 \begin{equation}\label{mvpure}
\begin{aligned}
I_f(|\psi\rangle,X)=V(|\psi\rangle,X)
\end{aligned}
\end{equation}
 for any pure state $\rho=|\psi\rangle\langle\psi|$, whereas
 \begin{equation}\label{mvmixed}
\begin{aligned}
 I_f(\rho,X)\leq V(\rho,X)
 \end{aligned}
\end{equation}
for a general mixed state $\rho$.

The metric-adjusted skew information and the variance possess several important properties \cite{CaiHansen2010,GuhnePRL2004,FrankHansen2008}:

(1) (Convexity and concavity) For any observable $X$, the metric-adjusted skew information is convex and the variance is concave, i.e.,

\begin{equation*}
\begin{aligned}
I_f\big(\sum\limits_ip_i\rho_i,X\big)&\leq\sum\limits_ip_iI_f(\rho_i,X), \\
V\big(\sum\limits_ip_i\rho_i,X\big)&\geq\sum\limits_ip_iV(\rho_i,X).
\end{aligned}
\end{equation*}
Here, $p_i\geq 0$ and $\sum\limits_ip_i=1$.

(2) (Additivity) For any $N$-partite product pure state $\bigotimes_{i=1}^N|\psi_i\rangle$ in the composite system $H_1\otimes
H_2\otimes\cdots\otimes H_N$, the metric-adjusted skew information and the variance satisfy
\begin{equation*}
\begin{aligned}
I_f\Big (\bigotimes\limits_{i=1}^N|\psi_i\rangle,\sum\limits_{i=1}^NX_i\Big )
=V\Big (\bigotimes\limits_{i=1}^N|\psi_i\rangle,\sum\limits_{i=1}^NX_i\Big
)=\sum\limits_{i=1}^NI_f(|\psi_i\rangle,X_i)=\sum\limits_{i=1}^NV(|\psi_i\rangle,X_i),
\end{aligned}
\end{equation*}
where $|\psi_i\rangle$ is a quantum state on the subsystem $H_i$, and $\sum\limits_{i=1}^NX_i=\sum\limits_{i=1}^N\mathbf{1}_1\otimes\cdots\otimes
\mathbf{1}_{i-1}\otimes X_i\otimes \mathbf{1}_{i+1}\otimes\cdots\otimes \mathbf{1}_N$. Here, $X_i$ is an observable acting on subsystem $H_i$, and $\mathbf{1}_j$ denotes the identity operator on subsystem $H_j$.

\section{$k$-nonstretchability criteria based on informationally complete $(s,t)$-POVMs}

Consider an $N$-partite quantum system $H_1\otimes H_2\otimes\cdots \otimes H_N$. For any subset $\omega \subseteq \{1,2,\ldots,N\}$, let $X_i$ be an operator on subsystem $H_i$, and define
$\mathbb{X}_\omega=\sum\limits_{i\in \omega}X_i=\sum\limits_{i\in \omega}\Big[X_i\otimes(\bigotimes\limits_{\substack{j\in\omega\\\textrm{ and
}j\neq i}}\mathbf{1}_j)\Big]$, which acts on the subsystem  $\bigotimes\limits_{i\in \omega}H_i$.

In particular, for $\omega=\{1,2,\ldots,N\}$, we write $\mathbb{X}=\mathbb{X}_{\{1,2,\ldots,N\}}$, namely,
$\mathbb{X}=\sum\limits_{i=1}^NX_i=\sum\limits_{i=1}^N\mathbf{1}_1\otimes\cdots\otimes \mathbf{1}_{i-1}\otimes X_i\otimes
\mathbf{1}_{i+1}\otimes\cdots\otimes \mathbf{1}_N=\mathbb{X}_{\{1,2,...,N\}}=\sum\limits_{i\in \{1,2,...,N\}}X_i$.

We define $I_{N+k}$ and $V_{N+k}$ as follows:
\begin{equation*}
\begin{aligned}
I_{N+k}
=N\Big[r^2t(\sqrt{t}+1)^2(d-1)-\dfrac{(d-1)(d+\Delta_{N+k})}{t(t-1)}\Big]+
\Big[\frac{s}{t}+r^2t(\sqrt{t}+1)^2(1-\frac{1}{d})\Big]\Gamma_{N+k}
\end{aligned}
\end{equation*}
and
\begin{equation*}
\begin{aligned}
V_{N+k}
= r^2t(\sqrt{t}+1)^2(d+1)N+\Big[\dfrac{s}{t}-r^2t(\sqrt{t}+1)^2(1+\dfrac{1}{d})-\dfrac{(d-1)(d^2+t^2\chi)}{dt(t-1)}\Big]\Gamma_{N+k}
\end{aligned}
\end{equation*}
respectively. Here
\begin{equation*}
\begin{aligned}
\Delta_{N+k}=\left\{\begin{array}{ll}1, &\textrm{ for }N+k\neq1,\\
  \dfrac{t^2\chi}{d} & \textrm{ for }N+k=1,
\end{array}
\right.
\end{aligned}
\end{equation*}
and
\begin{equation}\label{BB}
\begin{aligned}
\Gamma_{N+k}=\left\{\begin{array}{ll} \dfrac{(N+k)^2+4N-1}{4}, &\textrm{ if }N+k\textrm{ is odd,}\\
 34-k,& \textrm{ if }N+k= 10\textrm{ and }N\geq8,\\
  76-k,& \textrm{ if }N+k= 16\textrm{ and }N\geq12,\\
  \dfrac{(N+k)^2+2(N-k+4)}{4} & \textrm{ for other cases.}
\end{array}
\right.
\end{aligned}
\end{equation}
These special cases arise from the extremal values in Lemma~3 of Ref.~\cite{RenLiSmerziGessnerPRL2021}.

\emph{Theorem.}
In an $N$-partite quantum system $H_1\otimes H_2\otimes\cdots \otimes H_N$ with $\dim(H_i)=d$, let $\{A^{(uv)}:u=1,2,\ldots,s;\
v=1,2,\ldots,t \}$ be any informationally complete $(s,t)$-POVM on the Hilbert space $H_i$, and define $\mathbb{A}^{(uv)}=\sum_{i=1}^N A_i^{(uv)}$, where $A_i^{(uv)}$ denotes $A^{(uv)}$ acting on subsystem $H_i$.  If the quantum state $\rho$ is $k$-stretchable, then it satisfies the following two inequalities:

(I) (metric-adjusted skew information criterion)
\begin{equation}\label{MUMs-SI}
\begin{aligned}
\sum\limits_{u=1}^{s}\sum\limits_{v=1}^tI_f\Big(\rho, \mathbb{A}^{(uv)}\Big)
\leq I_{N+k},
\end{aligned}
\end{equation}

(II) (variance criterion)
\begin{equation}\label{MUMs-SII}
\begin{aligned}
\sum\limits_{u=1}^{s}\sum\limits_{v=1}^tV\Big(\rho, \mathbb{A}^{(uv)}\Big)
\geq V_{N+k}.
\end{aligned}
\end{equation}
If the quantum state $\rho$ violates either inequality~(\ref{MUMs-SI}) or~(\ref{MUMs-SII}), then $\rho$ is $k$-nonstretchable.

\emph{Proof.} We first prove the inequalities (\ref{MUMs-SI}) and (\ref{MUMs-SII}) for any $k$-stretchable pure state in a $d$-dimensional Hilbert space $|\psi\rangle \in H_1\otimes H_2\otimes\cdots \otimes H_N$. If an $N$-partite pure state $|\psi\rangle$ is $k$-stretchable (with $k+N\neq 1$), then there exists a partition $\gamma=\gamma_1|\cdots|\gamma_{|\gamma|}$ such that $|\psi\rangle=\bigotimes\limits_{i=1}^{|\gamma|}|\psi_{\gamma_i}\rangle$. Here, $\max\limits_{1\leq i\leq|\gamma|}n_{\gamma_i}-|\gamma|\leq k$ and $|\psi_{\gamma_i}\rangle$ is a state on the subsystem $\bigotimes\limits_{j\in \gamma_i}H_j$.

We first show that
\begin{align}
\sum\limits_{u=1}^{s}\sum\limits_{v=1}^tI_f\Big(|\psi\rangle, \mathbb{A}^{(uv)}\Big)
=&\sum\limits_{u=1}^{s}\sum\limits_{v=1}^t\sum\limits_{i=1}^{|\gamma|}I_f\Big(|\psi_{\gamma_i}\rangle,\mathbb{A}_{\gamma_i}^{(uv)}\Big)\label{V00}\\
=&\sum\limits_{u=1}^{s}\sum\limits_{v=1}^t\sum\limits_{i=1}^{|\gamma|}V\Big(|\psi_{\gamma_i}\rangle,\mathbb{A}_{\gamma_i}^{(uv)}\Big)\label{V0}\\
=&\sum\limits_{u=1}^{s}\sum\limits_{v=1}^t\sum\limits_{i=1}^{|\gamma|}\Big\{\textrm{Tr}\Big[\big(\mathbb{A}_{\gamma_i}^{(uv)}\big)^2|\psi_{\gamma_i}\rangle\langle\psi_{\gamma_i}|\Big]
-\Big[\textrm{Tr}\big(\mathbb{A}_{\gamma_i}^{(uv)}|\psi_{\gamma_i}\rangle\langle\psi_{\gamma_i}|\big)\Big]^2\Big\}\label{V1}\\
\leq&\sum\limits_{i=1}^{|\gamma|}\left[r^2t(\sqrt{t}+1)^2(d-1)n_{\gamma_i}+\Big(\frac{s}{t}+r^2t(\sqrt{t}+1)^2(1-\frac{1}{d})\Big)n_{\gamma_i}^2
-\dfrac{(d^2-1)n_{\gamma_i}}{t(t-1)}\right]\label{V2}\\
\leq& N\Big[r^2t(\sqrt{t}+1)^2(d-1)-\dfrac{(d^2-1)}{t(t-1)}\Big]+
\Big[\frac{s}{t}+r^2t(\sqrt{t}+1)^2(1-\frac{1}{d})\Big]\Big(\max\sum\limits_{i=1}^{|\gamma|}n_{\gamma_i}^2\Big)\label{V3}\\
= &I_{N+k}.\label{V4}
\end{align}
Here, Eq.~(\ref{V00}) follows from the additivity of the metric-adjusted skew information, and Eq.~(\ref{V0}) follows from the equivalence between the metric-adjusted skew information and the variance for pure states. Equation~(\ref{V1}) follows from the definition of the variance. Inequality (\ref{V2}) follows from inequalities (\ref{bound1}) and (\ref{bound10}):
\begin{equation}\label{bound1}
\begin{aligned}
\sum\limits_{u=1}^{s}\sum\limits_{v=1}^{t}\big(\mathbb{A}_{\gamma_i}^{(uv)}\big)^2 \leq
\Bigg[r^2t(\sqrt{t}+1)^2(d-1)n_{\gamma_i}+\Bigg(\frac{s}{t}+r^2t(\sqrt{t}+1)^2\Big(1-\frac{1}{d}\Big)\Bigg)n_{\gamma_i}^2
\Bigg]\bigotimes\limits_{j\in \gamma_i}\mathbf{1}_j ,
\end{aligned}
\end{equation}
and
\begin{equation}\label{bound10}
\begin{aligned}
\sum\limits_{u=1}^{s}\sum\limits_{v=1}^t\Big[\textrm{Tr}\big(\mathbb{A}_{\gamma_i}^{(uv)}|\psi_{\gamma_i}\rangle\langle\psi_{\gamma_i}|\big)\Big]^2
\geq&\dfrac{(d^2-1)n_{\gamma_i}}{t(t-1)}.
\end{aligned}
\end{equation}
The detailed proofs of these inequalities are provided in Appendix B. The maximum in Eq.~(\ref{V3}) is taken over all possible $k$-stretchable partitions. Equation (\ref{V4}) follows from Lemma 3 of Ref. \cite{RenLiSmerziGessnerPRL2021}.

Thus, we have shown that inequality~(\ref{MUMs-SI}) holds for any $k$-stretchable pure state $|\psi\rangle \in H_1\otimes H_2\otimes\cdots
\otimes H_N$ with $\dim(H_i)=d$ under the condition $k+N\neq1$.

Next, for the variance, we obtain
\begin{align}
\sum\limits_{u=1}^{s}\sum\limits_{v=1}^tV\Big(|\psi\rangle,\mathbb{A}^{(uv)}\Big)
=&\sum\limits_{i=1}^{|\gamma|}\Big\{\sum\limits_{u=1}^{s}\sum\limits_{v=1}^t\textrm{Tr}\Big[\big(\mathbb{A}_{\gamma_i}^{(uv)}\big)^2|\psi_{\gamma_i}\rangle\langle\psi_{\gamma_i}|\Big]
-\sum\limits_{u=1}^{s}\sum\limits_{v=1}^t\Big[\textrm{Tr}\big(\mathbb{A}_{\gamma_i}^{(uv)}|\psi_{\gamma_i}\rangle\langle\psi_{\gamma_i}|\big)\Big]^2\Big\}\label{VV0}\\
\geq &\sum\limits_{i=1}^{|\gamma|}
\left[r^2t(\sqrt{t}+1)^2(d+1)n_{\gamma_i}+\Big(\frac{s}{t}-r^2t(\sqrt{t}+1)^2(1+\frac{1}{d})\Big)n_{\gamma_i}^2
-\dfrac{(d-1)(d^2+t^2\chi)n_{\gamma_i}^2}{dt(t-1)}\right]\label{VV1}\\
\geq &r^2t(\sqrt{t}+1)^2(d+1)N+\Big[\frac{s}{t}-r^2t(\sqrt{t}+1)^2(1+\frac{1}{d})-\dfrac{(d-1)(d^2+t^2\chi)}{dt(t-1)}\Big]
\Big(\max\sum\limits_{i=1}^{|\gamma|}n_{\gamma_i}^2\Big)\label{VV2}\\
= &V_{N+k},\label{VV3}
\end{align}
where Eq.~(\ref{VV0}) follows from the additivity property and the definition of the variance, while inequality~(\ref{VV1}) is obtained from inequalities
(\ref{bound2}) and (\ref{bound3}):
\begin{equation}\label{bound2}
\begin{aligned}
\sum\limits_{u=1}^{s}\sum\limits_{v=1}^{t}\big(\mathbb{A}_{\gamma_i}^{(uv)}\big)^2 \geq
\Bigg[r^2t(\sqrt{t}+1)^2(d+1)n_{\gamma_i}+\Bigg(\frac{s}{t}-r^2t(\sqrt{t}+1)^2\Big(1+\frac{1}{d}\Big)\Bigg)n_{\gamma_i}^2
\Bigg]\bigotimes\limits_{j\in \gamma_i}\mathbf{1}_j ,
\end{aligned}
\end{equation}
and
\begin{equation}\label{bound3}
\begin{aligned}
\sum\limits_{u=1}^{s}\sum\limits_{v=1}^{t}\Big[\textrm{Tr}\big(\mathbb{A}_{\gamma_i}^{(uv)}|\psi_{\gamma_i}\rangle\langle\psi_{\gamma_i}|\big)\Big]^2
\leq\dfrac{(d-1)(d^2+t^2\chi)n_{\gamma_i}^2}{dt(t-1)}.
\end{aligned}
\end{equation}
The detailed proofs of these inequalities are also provided in Appendix B. The maximum in inequality (\ref{VV2}) is taken over all possible
$k$-stretchable partitions. Equation (\ref{VV3}) follows from Lemma 3 of Ref. \cite{RenLiSmerziGessnerPRL2021}.

Thus, we have proven that inequality~(\ref{MUMs-SII}) holds for any $k$-stretchable pure state $|\psi\rangle \in H_1\otimes H_2\otimes\cdots
\otimes H_N$ with $\dim(H_i)=d$ under the condition $k+N\neq1$.

If an $N$-partite pure state $|\psi\rangle$ is $(1-N)$-stretchable (i.e., $k+N=1$), then it can be written as
$|\psi\rangle=\bigotimes\limits_{i=1}^N|\psi_i\rangle$, where $|\psi_i\rangle$ is a state on the subsystem $H_i$. Using Eq.~(\ref{mvpure}), the additivity property, and the definition of the variance, we have
\begin{align}
&\sum\limits_{u=1}^{s}\sum\limits_{v=1}^{t}I_f\Big(|\psi\rangle,\mathbb{A}^{(uv)}\Big)\nonumber\\
=&\sum\limits_{u=1}^{s}\sum\limits_{v=1}^{t}\sum\limits_{i=1}^{N}\Big\{\textrm{Tr}\Big[\big(A_{i}^{(uv)}\big)^2|\psi_{i}\rangle\langle\psi_{i}|\Big]
-\Big[\textrm{Tr}\big(A_{i}^{(uv)}|\psi_{i}\rangle\langle\psi_{i}|\big)\Big]^2\Big\}\nonumber\\
=&N\Big[\frac{s}{t}+r^2t(\sqrt{t}+1)^2(d-\frac{1}{d})-\frac{d(t^2\chi-d)+d^3-t^2\chi}{dt(t-1)}\Big],\label{bound}
\end{align}
where Eq. (\ref{bound}) follows from Eqs. (\ref{bound0}) and (\ref{bound00}). Thus, inequalities (\ref{MUMs-SI}) and (\ref{MUMs-SII}) hold for
any $(1-N)$-stretchable pure state. It follows from Eq.~(\ref{bound}) that, for $(1-N)$-stretchable pure states, equalities hold in (\ref{MUMs-SI}) and (\ref{MUMs-SII}).
Consequently, any pure state that does not satisfy Eq.~(\ref{bound}) is $(1-N)$-nonstretchable.

So far, we have shown that inequalities (\ref{MUMs-SI}) and (\ref{MUMs-SII}) hold for any $k$-stretchable pure state.

Next, consider any $N$-partite $k$-stretchable mixed state $\rho=\sum\limits_ip_i|\psi_i\rangle\langle\psi_i|$ on the Hilbert space $H_1\otimes
H_2\otimes\cdots \otimes H_N$ with $\textrm{dim}(H_i)=d$, where each pure state $|\psi_i\rangle$ is $k$-stretchable. By the convexity of the metric-adjusted skew information and the validity of inequality~(\ref{MUMs-SI}) for pure states, inequality~(\ref{MUMs-SI}) also holds for any $k$-stretchable mixed state. Similarly, by the concavity of the variance and the validity of inequality~(\ref{MUMs-SII}) for pure states, inequality~(\ref{MUMs-SII}) also holds for any $k$-stretchable mixed state.

Thus, we have proven that both inequalities~(\ref{MUMs-SI}) and~(\ref{MUMs-SII}) are valid for any $k$-stretchable mixed state. This completes the proof of the theorem.

This theorem provides a method for identifying $k$-nonstretchable states. For a given quantum state $\rho$, if it violates either of the two informationally complete $(s,t)$-POVM-based criteria, then the state is $k$-nonstretchable, meaning that $\rho\notin C_k$, that is, $\rho\in C_{N-1}\setminus C_k$. Consequently, the theorem provides a systematic method for classifying multipartite entanglement. A concrete illustration of this classification is provided in Fig.~2.
\begin{figure}[h]
\centering
\includegraphics[width=0.5\textwidth]{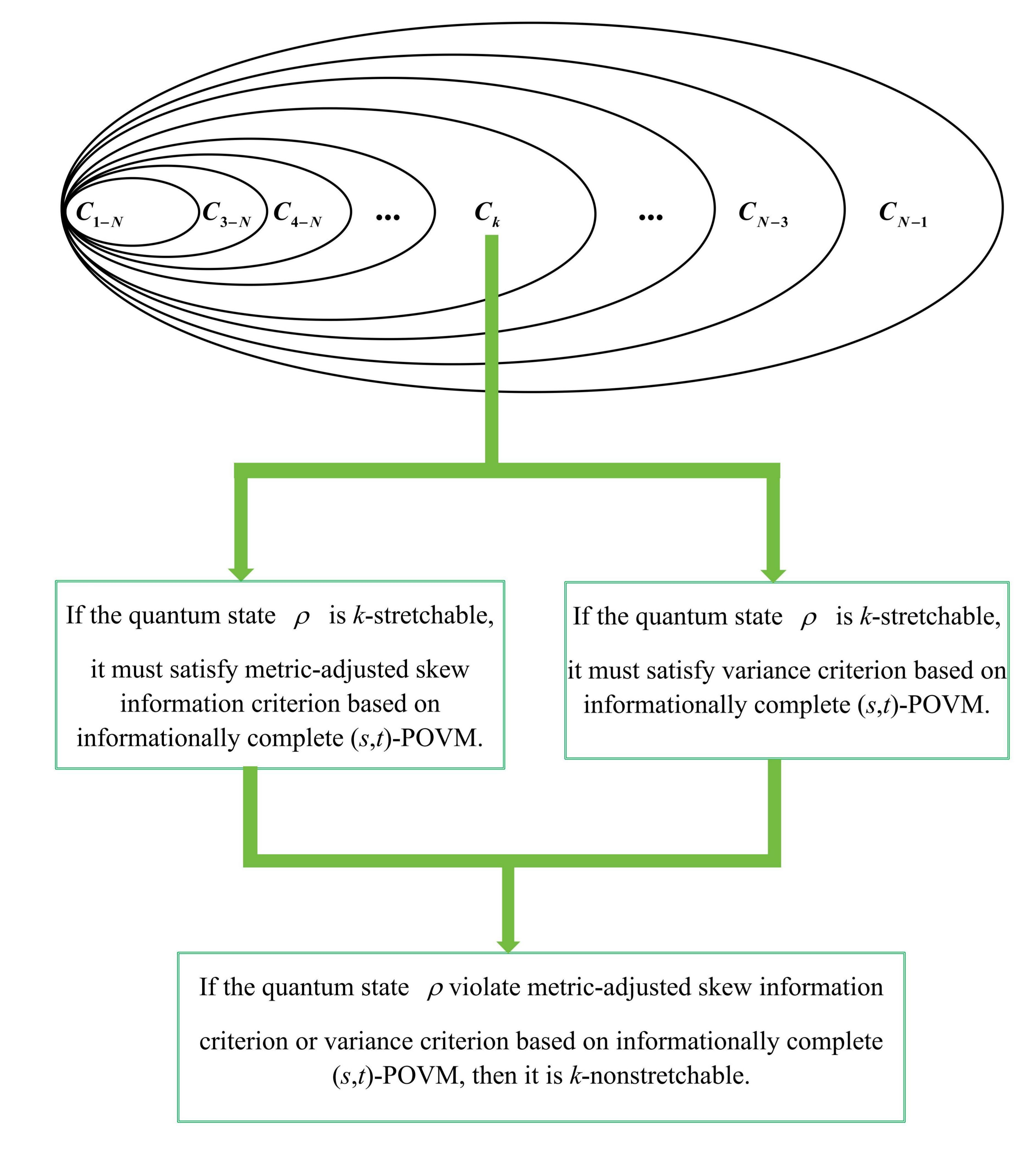}
\caption{In an $N$-partite quantum system, the set $C_k$ represents the collection of all $k$-stretchable states, where $k=1-N,3-N,4-N,\cdots,N-3,N-1$. The set $C_{N-1}$ is the set of all quantum states. These sets are convex and satisfy the inclusion relation $C_{1-N}\subset C_{3-N}\subset C_{4-N}\subset\cdots\subset C_{k}\subset C_{N-3}\subset C_{N-1}$. Moreover, $C_{N-1}\setminus C_k$ constitutes the set of all $k$-nonstretchable states. Any $k$-stretchable state must satisfy the two informationally complete $(s,t)$-POVM-based criteria, namely the metric-adjusted skew information criterion and the variance criterion. In other words, if a quantum state violates either of the two informationally complete $(s,t)$-POVM-based criteria, then the state is $k$-nonstretchable, meaning that $\rho\in C_{N-1}\setminus C_k$. Thus, the two informationally complete $(s,t)$-POVM-based criteria offer a method for classifying multipartite entanglement.}
\end{figure}

Since both criteria are based on informationally complete $(s,t)$-POVMs and involve observables, they are particularly suitable for experimental implementation. The violation of either criterion indicates that the quantum state is $k$-nonstretchable. In particular, once a state is identified as $(1-N)$-nonstretchable, it must be entangled. This, in turn, provides a practical approach for detecting entanglement.

\section{Illustrations}

In view of the relations between the variance and the skew information given in Eqs.~(\ref{mvpure}) and~(\ref{mvmixed}), we discuss the relationship between the variance criterion and the metric-adjusted skew information criterion for $k$-nonstretchability. Since $I_{N+k}\geq V_{N+k}$, we obtain the following conclusions:

(1) If $\sum\limits_{u=1}^{s}\sum\limits_{v=1}^tI_f\Big(\rho, \mathbb{A}^{(uv)}\Big)> I_{N+k}$, then $\rho$ is $k$-nonstretchable and can be detected by the $k$-nonstretchability criterion based on the metric-adjusted skew information.

(2) If $\sum\limits_{u=1}^{s}\sum\limits_{v=1}^tV\Big(\rho, \mathbb{A}^{(uv)}\Big)< V_{N+k}$, then $\rho$ is $k$-nonstretchable and can be detected by the $k$-nonstretchability criterion based on the variance.

(3) If either $V_{N+k}\leq\sum\limits_{u=1}^{s}\sum\limits_{v=1}^tI_f\Big(\rho, \mathbb{A}^{(uv)}\Big)\leq I_{N+k}$ or $\sum\limits_{u=1}^{s}\sum\limits_{v=1}^tI_f\Big(\rho, \mathbb{A}^{(uv)}\Big)\leq V_{N+k}\leq\sum\limits_{u=1}^{s}\sum\limits_{v=1}^tV\Big(\rho, \mathbb{A}^{(uv)}\Big)$, then neither criterion can detect the $k$-nonstretchability of $\rho$.

To demonstrate the effectiveness of the proposed criteria, we present illustrative examples and compare them with two established criteria, namely Eq.~(6) of Ref.~\cite{RenLiSmerziGessnerPRL2021}, based on Young diagrams, and Eq.~(3) of Ref.~\cite{HongGaoYanPLA2021}, based on the operator inequality. For convenience, we refer to these two criteria as the Young-diagram criterion and the operator-inequality criterion, respectively.

\emph{Example 1.} Consider the $N$-qutrit state $|G\rangle=\dfrac{|0\rangle^{\otimes N}+|1\rangle^{\otimes N}+|2\rangle^{\otimes N}}{\sqrt{3}}$ mixed with the maximally mixed state, namely,
\begin{equation*}
\begin{aligned}
\rho_1(p)=p|G\rangle\langle G|+\dfrac{1-p}{3^N}\mathbf{1},  \quad 0\leq p\leq1.
\end{aligned}
\end{equation*}
For $d=3$, we construct an informationally complete $(1,9)$-POVM using the generalized Gell-Mann operator basis \cite{BertlmannKrammer2008}. This construction yields the range $0.0121\leq r\leq0.0129$. In this example, we choose $r=0.0129$. A direct calculation shows that the variance criterion does not detect any $(3-N)$-nonstretchable states in this family, whereas the metric-adjusted skew information criterion detects certain $(3-N)$-nonstretchable states when $N\geq6$. Specifically, according to the metric-adjusted skew information criterion, $\rho_1(p)$ is $(3-N)$-nonstretchable for $N=10,20,30,40,50$ when $0.5156<p\leq1$, $0.2549<p\leq1$, $0.1692<p\leq1$, $0.1266<p\leq1$, and $0.1011<p\leq1$, respectively. For this family of states, the Young-diagram criterion \cite{RenLiSmerziGessnerPRL2021} cannot detect any $(3-N)$-nonstretchable states because it applies only when $\dim(H_i)=2$. Hence, for the family of states $\rho_1(p)$, the metric-adjusted skew information criterion performs better than both the variance criterion and the Young-diagram criterion \cite{RenLiSmerziGessnerPRL2021} in identifying $(3-N)$-nonstretchable states. These results are summarized in Table~I.

\begin{table}
\caption{\label{tab:table1}
Parameter thresholds for detecting $(3-N)$-nonstretchable states in the family $\rho_1(p)$. The symbol ``\textbackslash'' indicates that neither the variance criterion nor the Young-diagram criterion \cite{RenLiSmerziGessnerPRL2021} can detect $(3-N)$-nonstretchable states in this family. The parameter ranges detected by the metric-adjusted skew information criterion for $N=10,20,30,40,50$ are $0.5156<p\leq1$, $0.2549<p\leq1$, $0.1692<p\leq1$, $0.1266<p\leq1$, and $0.1011<p\leq1$, respectively.}
\begin{ruledtabular}
\begin{tabular}{cccccc}
$N$ &10&20&30&40&50\\
\hline
Metric-adjusted skew information criterion
& $p>0.5156$ & $p>0.2549$ & $p>0.1692$ & $p>0.1266$ & $p>0.1011$  \\

Variance criterion
& $\backslash$ & $\backslash$ & $\backslash$ & $\backslash$ & $\backslash$  \\

Young-diagram criterion \cite{RenLiSmerziGessnerPRL2021}  & $\backslash$ & $\backslash$ & $\backslash$ & $\backslash$ & $\backslash$
\end{tabular}
\end{ruledtabular}
\end{table}

Furthermore, we observe that the metric-adjusted skew information criterion can identify entangled states, i.e., $(1-N)$-nonstretchable states. Specifically, $\rho_1(p)$ is entangled for $N=10,20,30,40,50$ when $0.2767<p\leq1$, $0.1595<p\leq1$, $0.1120<p\leq1$, $0.0863<p\leq1$, and $0.0702<p\leq1$, respectively. However, for $N=10,20,30,40,50$, the operator-inequality criterion \cite{HongGaoYanPLA2021} cannot detect any entangled states for either of the following choices: $|\psi\rangle=|0\rangle^{\otimes N}$, $\{\omega_1,\cdots,\omega_1\}=\{1\}$ or
$|\psi\rangle=\big(\frac{|0\rangle+|1\rangle}{\sqrt{2}}\big)^{\otimes N}$,
$\{\omega_1,\cdots,\omega_1\}=\{\frac{|0\rangle-|1\rangle}{\sqrt{2}}\}$. These results are summarized in Table~II.
\begin{table}
\caption{\label{tab:table2}
Parameter thresholds for detecting entangled states in the family $\rho_1(p)$. The symbol ``\textbackslash'' indicates that the operator-inequality criterion \cite{HongGaoYanPLA2021} cannot detect any entangled states in the family $\rho_1(p)$.
The parameter range of entangled states for $N=10,20,30,40,50$ detected by metric-adjusted skew information criterion is
 $0.2767<p\leq1$, $0.1595<p\leq1$, $0.1120<p\leq1$, $0.0863<p\leq1$, and $0.0702<p\leq1$, respectively.
}
\begin{ruledtabular}
\begin{tabular}{cccccc}
$N$ &10&20&30&40&50\\
\hline
Metric-adjusted skew information criterion & $p>$0.2767 & $p>$0.1595 & $p>$0.1120 & $p>$0.0863 & $p>$0.0702  \\
Operator-inequality criterion \cite{HongGaoYanPLA2021}  & $\backslash$ & $\backslash$ & $\backslash$ & $\backslash$ & $\backslash$
\end{tabular}
\end{ruledtabular}
\end{table}

\emph{Example 2.} To further illustrate the applicability of our criteria, we consider another important family of states, namely the $N$-partite state $\rho_2(p)$ defined as a mixture of the antisymmetric state $|S_N\rangle$ and white noise:
\begin{equation*}
\begin{aligned}
\rho_2(p)=p|S_N\rangle\langle S_N|+\dfrac{1-p}{N^N}\mathbf{1},  \quad 0\leq p\leq1,
\end{aligned}
\end{equation*}
where $|S_N\rangle=\dfrac{1}{\sqrt{N!}}\sum\limits_\sigma(-1)^{\textrm{sgn}(\sigma)}|\sigma\rangle$, $\textrm{sgn}(\sigma)$ denotes the signature of the permutation $\sigma$ of the sequence $01\cdots(N-1)$, and the sum is taken over all permutations.

In this example, we construct an informationally complete $(1,N^2)$-POVM using the generalized Gell-Mann operator basis \cite{BertlmannKrammer2008}. We find that the metric-adjusted skew information criterion does not detect any $(3-N)$-nonstretchable states in this family, whereas the variance criterion detects certain $(3-N)$-nonstretchable states. In particular, according to the variance criterion, $\rho_2(p)$ is $(3-N)$-nonstretchable when $$p>\dfrac{N^3(N+1)(N+3)r^2+N-1}{N^3(N+1)^2(N-1)r^2+N-1}.$$ For this family of states, the Young-diagram criterion \cite{RenLiSmerziGessnerPRL2021} is again ineffective because it applies only when $\dim(H_i)=2$. Therefore, the variance criterion performs better than both the metric-adjusted skew information criterion and the Young-diagram criterion \cite{RenLiSmerziGessnerPRL2021} in detecting $(3-N)$-nonstretchable states. Furthermore, we find that the variance criterion can identify entangled states, i.e., $(1-N)$-nonstretchable states. Specifically, $\rho_2(p)$ is entangled when $$p>\dfrac{N^3(N+1)r^2+1}{N^3(N+1)^2r^2+1}.$$ However, the operator-inequality criterion \cite{HongGaoYanPLA2021} cannot detect any entangled states for either choice of the test vectors, namely $|\psi\rangle=|0\rangle^{\otimes N}$,
 $\{\omega_1,\cdots,\omega_1\}=\{1\}$ or $|\psi\rangle=\big(\frac{|0\rangle+|1\rangle}{\sqrt{2}}\big)^{\otimes N}$,
 $\{\omega_1,\cdots,\omega_1\}=\{\frac{|0\rangle-|1\rangle}{\sqrt{2}}\}$. Therefore, for $\rho_2(p)$, the variance criterion exhibits stronger entanglement detection power than the operator-inequality criterion \cite{HongGaoYanPLA2021}.

Collectively, these results demonstrate that the proposed metric-adjusted skew information criterion and variance criterion exhibit complementary strengths in identifying $(3-N)$-nonstretchable states across different classes of quantum states. Their combined application substantially enhances the detection power, thereby surpassing the Young-diagram criterion \cite{RenLiSmerziGessnerPRL2021} for $\rho_1(p)$ and $\rho_2(p)$. In addition, the metric-adjusted skew information criterion and the variance criterion exhibit stronger entanglement detection power than the operator-inequality criterion \cite{HongGaoYanPLA2021} for $\rho_1(p)$ and $\rho_2(p)$.
\section{Discussion}

The detection of $k$-nonstretchability reveals intricate structures of multipartite entanglement and provides a framework for characterizing entanglement hierarchies. In this work, we have investigated the characterization of $k$-nonstretchability using informationally complete $(s,t)$-POVMs. By employing informationally complete $(s,t)$-POVMs as observables in both metric-adjusted skew information and variance, we have derived two criteria for detecting $k$-nonstretchable states. These criteria are directly applicable and can detect specific classes of $k$-nonstretchable states. Their detection effectiveness has been demonstrated through explicit examples, which show in particular that they can identify $k$-nonstretchable states that are not detected by the other criteria considered here. However, neither criterion is necessary and sufficient; therefore, they cannot detect all $k$-nonstretchable states. Moreover, $k$-nonstretchability coincides with the conventional notion of entanglement only in the special case $k=1-N$. In general, $k$-nonstretchability is distinct from $k$-partite entanglement and $k$-nonseparability, including genuine multipartite entanglement. A deeper investigation of the relationships among these notions is therefore required.

Given the substantial progress in constructing entanglement detection criteria, a natural direction for future research is to explore whether general methods for identifying $k$-nonstretchability can be systematically developed based directly on informationally complete $(s,t)$-POVMs. Such methods would broaden the applicability of these approaches to multipartite entanglement detection.

\subsection*{Lead contact}

Further information and requests for resources should be directed to the lead contact Limin Gao (gaoliminabc@163.com).

\section*{Data availability statement}

No data was used for the research described in the article.

\section*{Acknowledgments}

This work was supported by the Hebei Natural Science Foundation of China under Grant No.~A2025403008, the National Pre-research Funds of Hebei GEO University under Grant No.~KY2025YB15, and the Hengshui University High-level Talent Research Fund under Grant No.~2022GC11.

\appendix

\section{Proof of Eq.~(\ref{bound0})}

It was shown in Ref.~\cite{SiudzinskaPRA2022} that any informationally complete $(s,t)$-POVM can be obtained using the method described in Sec.~II B. Hence, in a Hilbert space $H$ with $\dim(H)=d$, let $\{A^{(uv)}:u=1,2,\cdots, s;v=1,2,\cdots,t\}$ be an informationally complete $(s,t)$-POVM with parameter $r$, and let $B^{(uv)}=\dfrac{A^{(uv)}-\frac{\mathbf{1}}{t}}{r}$ for $u=1,2,\cdots,s$ and $v=1,2,\cdots,t$.
Then the set $\{\dfrac{\mathbf{1}}{\sqrt{d}}, C^{(uv)}:u=1,2,\cdots, s;v=1,2,\cdots,t-1\}$, where
\begin{equation}\label{appendixLOO}
\begin{aligned}
C^{(uv)}=\dfrac{B^{(ut)}}{\sqrt{t}(\sqrt{t}+1)^2}-\dfrac{B^{(uv)}}{\sqrt{t}(\sqrt{t}+1)},
\end{aligned}
\end{equation}
constitutes an orthonormal Hermitian operator basis. It follows that
\begin{align}
\sum\limits_{u=1}^{s}\sum\limits_{v=1}^t\big(A^{(uv)}\big)^2
=&\sum\limits_{u=1}^{s}\sum\limits_{v=1}^t\Big(\dfrac{\mathbf{1}}{t}+rB^{(uv)}\Big)^2\label{appendixLOO1}\\
=&\sum\limits_{u=1}^{s}\sum\limits_{v=1}^t\dfrac{\mathbf{1}}{t^2}+\dfrac{2r}{t}\sum\limits_{u=1}^{s}\sum\limits_{v=1}^tB^{(uv)}
+r^2\sum\limits_{u=1}^{s}\sum\limits_{v=1}^t\Big(B^{(uv)}\Big)^2\nonumber\\
=&\dfrac{s}{t}\mathbf{1}+\dfrac{2r}{t}\sum\limits_{u=1}^{s}\sum\limits_{v=1}^{t-1}\Big[C^{(u)}-\sqrt{t}(\sqrt{t}+1)C^{(uv)}\Big]
+\dfrac{2r}{t}\sum\limits_{u=1}^{s}(\sqrt{t}+1)C^{(u)}+r^2\sum\limits_{u=1}^{s}\sum\limits_{v=1}^t(B^{(uv)})^2\label{appendixLOO2}\\
=&\dfrac{s}{t}\mathbf{1}+2r(1+\dfrac{1}{\sqrt{t}})\sum\limits_{u=1}^{s}C^{(u)}
-2r(1+\dfrac{1}{\sqrt{t}})\sum\limits_{u=1}^{s}\sum\limits_{v=1}^{t-1}C^{(uv)}+r^2\sum\limits_{u=1}^{s}\sum\limits_{v=1}^t\Big(B^{(uv)}\Big)^2\nonumber\\
=&\dfrac{s}{t}\mathbf{1}+r^2\sum\limits_{u=1}^{s}\sum\limits_{v=1}^t\Big(B^{(uv)}\Big)^2\label{appendixLOO3}\\
=&\dfrac{s}{t}\mathbf{1}+r^2\sum\limits_{u=1}^{s}\sum\limits_{v=1}^{t-1}\Big[C^{(u)}-\sqrt{t}(\sqrt{t}+1)C^{(uv)}\Big]^2
+r^2\sum\limits_{u=1}^{s}\Big[(\sqrt{t}+1)C^{(u)}\Big]^2\label{appendixLOO4}\\
=&\dfrac{s}{t}\mathbf{1}+r^2\sum\limits_{u=1}^{s}\sum\limits_{v=1}^{t-1}\Big[C^{(u)}\Big]^2
-r^2\sqrt{t}(\sqrt{t}+1)\sum\limits_{u=1}^{s}\sum\limits_{v=1}^{t-1}C^{(u)}C^{(uv)}-r^2\sqrt{t}(\sqrt{t}+1)\sum\limits_{u=1}^{s}\sum\limits_{v=1}^{t-1}C^{(uv)}C^{(u)}\nonumber\\
&+r^2t(\sqrt{t}+1)^2\sum\limits_{u=1}^{s}\sum\limits_{v=1}^{t-1}\Big[C^{(uv)}\Big]^2+r^2(\sqrt{t}+1)^2\sum\limits_{u=1}^{s}\Big[C^{(u)}\Big]^2\nonumber\\
=&\dfrac{s}{t}\mathbf{1}+r^2t(\sqrt{t}+1)^2\sum\limits_{u=1}^{s}\sum\limits_{v=1}^{t-1}\Big[C^{(uv)}\Big]^2\label{appendixLOO5}\\
=&\Big[\frac{s}{t}+r^2t(\sqrt{t}+1)^2(d-\frac{1}{d})\Big] \mathbf{1}.\label{appendixLOO6}
\end{align}
Here, Eq.~(\ref{appendixLOO1}) follows from Eq.~(\ref{symmetricmeasurements2}). Equations~(\ref{appendixLOO2}) and~(\ref{appendixLOO4}) follow from Eq.~(\ref{symmetricmeasurements1}). Equations~(\ref{appendixLOO3}) and~(\ref{appendixLOO5}) follow from the relation $C^{(u)}=\sum_{v=1}^{t-1}C^{(uv)}$. Finally, Eq.~(\ref{appendixLOO6}) follows from the fact that the set $\Big\{\dfrac{\mathbf{1}}{\sqrt{d}}, C^{(uv)}:u=1,2,\cdots, s;v=1,2,\cdots,t-1\Big\}$ constitutes an orthonormal Hermitian operator basis.

\section{Proofs of inequalities~(\ref{bound1}),~(\ref{bound10}),~(\ref{bound2}), and~(\ref{bound3})}

When $n_{\gamma_i}=1$, we obtain
\begin{equation}\label{bound11}
\begin{aligned}
\sum\limits_{u=1}^{s}\sum\limits_{v=1}^t\big(A_{\gamma_i}^{(uv)}\big)^2=\Big[\frac{s}{t}+r^2t(\sqrt{t}+1)^2(d-\frac{1}{d})\Big] \mathbf{1},
\end{aligned}
\end{equation}
which follows from Eq.~(\ref{appendixLOO6}). When $n_{\gamma_i}>1$, applying Eq.~(\ref{bound11}) to each subsystem $H_j$, we obtain
\begin{equation}\label{bound12}
\begin{aligned}
\sum\limits_{u=1}^{s}\sum\limits_{v=1}^t\sum\limits_{j\in \gamma_i}\big(A_{j}^{(uv)}\big)^2
=n_{\gamma_i}\Big[\frac{s}{t}+r^2t(\sqrt{t}+1)^2(d-\frac{1}{d})\Big] \bigotimes\limits_{j\in \gamma_i}\mathbf{1}_j.
\end{aligned}
\end{equation}
In addition, for $j\neq j'$, we have
\begin{align}
&\sum\limits_{u=1}^{s}\sum\limits_{v=1}^tA^{(uv)}_j\otimes A^{(uv)}_{j'}\label{}\nonumber\\
=&\sum\limits_{u=1}^{s}\sum\limits_{v=1}^t\Big(\dfrac{\mathbf{1}}{t}+rB^{(uv)}\Big)\otimes\Big(\dfrac{\mathbf{1}}{t}+rB^{(uv)}\Big)\label{appendixLOO11}\\
=&\frac{s}{t}\mathbf{1}\otimes\mathbf{1}
+\frac{r}{t}\sum\limits_{u=1}^{s}\sum\limits_{v=1}^{t-1}\mathbf{1}\otimes\Big[C^{(u)}-\sqrt{t}(\sqrt{t}+1) C^{(uv)}\Big]
+\frac{r}{t}\sum\limits_{u=1}^{s}\mathbf{1}\otimes\Big[(\sqrt{t}+1) C^{(u)}\Big]\label{appendixLOO12}\\
+&\frac{r}{t}\sum\limits_{u=1}^{s}\sum\limits_{v=1}^{t-1}\Big[C^{(u)}-\sqrt{t}(\sqrt{t}+1) C^{(uv)}\Big]\otimes\mathbf{1}
+\frac{r}{t}\sum\limits_{u=1}^{s}\Big[(\sqrt{t}+1) C^{(u)}\Big]\otimes\mathbf{1}\label{appendixLOO13}\\
+&r^2\sum\limits_{u=1}^{s}\sum\limits_{v=1}^{t-1}\Big[C^{(u)}-\sqrt{t}(\sqrt{t}+1) C^{(uv)}\Big]\otimes\Big[C^{(u)}-\sqrt{t}(\sqrt{t}+1)
C^{(uv)}\Big]
+r^2\sum\limits_{u=1}^{s}\Big[(\sqrt{t}+1) C^{(u)}\Big]\otimes\Big[(\sqrt{t}+1) C^{(u)}\Big]\label{appendixLOO14}\\
=&\frac{s}{t}\mathbf{1}\otimes\mathbf{1}+r^2t(\sqrt{t}+1)^2\sum\limits_{u=1}^{s}\sum\limits_{v=1}^{t-1} C^{(uv)}\otimes
C^{(uv)}.\label{appendixLOO15}
\end{align}
Here, Eq.~(\ref{appendixLOO11}) follows from Eq.~(\ref{symmetricmeasurements2}), while Eqs.~(\ref{appendixLOO12})--(\ref{appendixLOO14}) follow from Eq.~(\ref{symmetricmeasurements1}). Equation~(\ref{appendixLOO15}) follows from the relation $C^{(u)}=\sum_{v=1}^{t-1}C^{(uv)}$. Since the set $\Big\{\dfrac{\mathbf{1}}{\sqrt{d}}, C^{(uv)}:u=1,2,\cdots, s;v=1,2,\cdots,t-1\Big\}$ constitutes an orthonormal Hermitian operator basis, one has
$$-(1+\frac{1}{d})\mathbf{1}\otimes\mathbf{1}\leq\sum\limits_{u=1}^{s}\sum\limits_{v=1}^{t-1} C^{(uv)}\otimes
C^{(uv)}\leq(1-\frac{1}{d})\mathbf{1}\otimes\mathbf{1}.$$
Thus, we obtain
\begin{align}
\Big[\frac{s}{t}-r^2t(\sqrt{t}+1)^2(1+\frac{1}{d})\Big]\mathbf{1}\otimes\mathbf{1}\leq\sum\limits_{u=1}^{s}\sum\limits_{v=1}^tA^{(uv)}_j\otimes
A^{(uv)}_{j'}
\leq\Big[\frac{s}{t}+r^2t(\sqrt{t}+1)^2(1-\frac{1}{d})\Big]\mathbf{1}\otimes\mathbf{1}.\label{}\nonumber
\end{align}
Hence, when $n_{\gamma_i}>1$, one has
\begin{equation*}
\begin{aligned}
&\sum\limits_{u=1}^{s}\sum\limits_{v=1}^{t}\big(\mathbb{A}_{\gamma_i}^{(uv)}\big)^2\\
=&\sum\limits_{u=1}^{s}\sum\limits_{v=1}^{t}\sum\limits_{j\in
\gamma_i}(A_{j}^{(uv)})^2+2\sum\limits_{u=1}^{s}\sum\limits_{v=1}^{t}\sum\limits_{\substack{j, j'\in \gamma_i\\\textrm{ and }j<
j'}}\Big[A_{j}^{(uv)}\otimes A_{j'}^{(uv)}\otimes(\bigotimes\limits_{\substack{l\in  \gamma_i\textrm{ and }\\ l\neq j,l\neq
j'}}\mathbf{1}_l)\Big]\\
\leq&n_{\gamma_i}\Big[\frac{s}{t}+r^2t(\sqrt{t}+1)^2(d-\frac{1}{d})\Big] \bigotimes\limits_{j\in
\gamma_i}\mathbf{1}_j+n_{\gamma_i}(n_{\gamma_i}-1)\Big[\frac{s}{t}+r^2t(\sqrt{t}+1)^2(1-\frac{1}{d})\Big]\bigotimes\limits_{j\in
\gamma_i}\mathbf{1}_j\\
= &\left[r^2t(\sqrt{t}+1)^2(d-1)n_{\gamma_i}+\Big(\frac{s}{t}+r^2t(\sqrt{t}+1)^2(1-\frac{1}{d})\Big)n_{\gamma_i}^2 \right]\bigotimes\limits_{j\in
\gamma_i}\mathbf{1}_j,
\end{aligned}
\end{equation*}
and
\begin{equation*}
\begin{aligned}
&\sum\limits_{u=1}^{s}\sum\limits_{v=1}^{t}\big(\mathbb{A}_{\gamma_i}^{(uv)}\big)^2\\
=&\sum\limits_{u=1}^{s}\sum\limits_{v=1}^{t}\sum\limits_{j\in
\gamma_i}\big(A_{j}^{(uv)}\big)^2+2\sum\limits_{u=1}^{s}\sum\limits_{v=1}^{t}\sum\limits_{\substack{j, j'\in \gamma_i\\\textrm{ and }j<
j'}}\Big[A_{j}^{(uv)}\otimes A_{j'}^{(uv)}\otimes(\bigotimes\limits_{\substack{l\in  \gamma_i\textrm{ and }\\ l\neq j,l\neq
j'}}\mathbf{1}_l)\Big]\\
\geq&n_{\gamma_i}\Big[\frac{s}{t}+r^2t(\sqrt{t}+1)^2(d-\frac{1}{d})\Big] \bigotimes\limits_{j\in
\gamma_i}\mathbf{1}_j+n_{\gamma_i}(n_{\gamma_i}-1)\Big[\frac{s}{t}-r^2t(\sqrt{t}+1)^2(1+\frac{1}{d})\Big]\bigotimes\limits_{j\in
\gamma_i}\mathbf{1}_j\\
= &\left[r^2t(\sqrt{t}+1)^2(d+1)n_{\gamma_i}+\Big(\frac{s}{t}-r^2t(\sqrt{t}+1)^2(1+\frac{1}{d})\Big)n_{\gamma_i}^2 \right]\bigotimes\limits_{j\in
\gamma_i}\mathbf{1}_j,
\end{aligned}
\end{equation*}
In summary, we obtain the following inequalities:
\begin{equation*}
\begin{aligned}
\left[r^2t(\sqrt{t}+1)^2(d+1)n_{\gamma_i}+\Big(\frac{s}{t}-r^2t(\sqrt{t}+1)^2(1+\frac{1}{d})\Big)n_{\gamma_i}^2 \right]\bigotimes\limits_{j\in
\gamma_i}\mathbf{1}_j
\leq\sum\limits_{u=1}^{s}\sum\limits_{v=1}^{t}\big(\mathbb{A}_{\gamma_i}^{(uv)}\big)^2,
\end{aligned}
\end{equation*}
and
 \begin{equation*}
\begin{aligned}
\sum\limits_{u=1}^{s}\sum\limits_{v=1}^{t}\big(\mathbb{A}_{\gamma_i}^{(uv)}\big)^2
\leq \left[r^2t(\sqrt{t}+1)^2(d-1)n_{\gamma_i}+\Big(\frac{s}{t}+r^2t(\sqrt{t}+1)^2(1-\frac{1}{d})\Big)n_{\gamma_i}^2
\right]\bigotimes\limits_{j\in \gamma_i}\mathbf{1}_j.
\end{aligned}
\end{equation*}
Both inequalities reduce to equalities when $n_{\gamma_i}=1$. These two inequalities correspond precisely to inequalities~(\ref{bound2}) and~(\ref{bound1}), respectively.
When $n_{\gamma_i}=1$, the following identity holds \cite{SiudzinskaPRA2022}
 \begin{equation*}
\begin{aligned}
\sum\limits_{u=1}^{s}\sum\limits_{v=1}^t\Big[\textrm{Tr}\big(A^{(uv)}\rho\big)\Big]^2=&\frac{d(t^2\chi-d)\textrm{Tr}(\rho^2)+d^3-t^2\chi}{dt(t-1)}.
\end{aligned}
\end{equation*}
When $n_{\gamma_i}>1$, we have
\begin{align}
&\sum\limits_{u=1}^{s}\sum\limits_{v=1}^{t}\Big[\textrm{Tr}\big(\mathbb{A}_{\gamma_i}^{(uv)}|\psi_{\gamma_i}\rangle\langle\psi_{\gamma_i}|\big)\Big]^2\label{}\nonumber\\
=&\sum\limits_{u=1}^{s}\sum\limits_{v=1}^{t}\Big[\sum\limits_{j\in \gamma_i}\textrm{Tr}\big(A_{j}^{(uv)}\rho_j\big)\Big]^2\label{bound101}\\
\geq& \sum\limits_{u=1}^{s}\sum\limits_{v=1}^{t}\sum\limits_{j\in \gamma_i}\Big[\textrm{Tr}\big(A_{j}^{(uv)}\rho_j\big)\Big]^2\label{bound102}\\
=&\sum\limits_{j\in \gamma_i}\frac{d(t^2\chi-d)\textrm{Tr}(\rho_j^2)+d^3-t^2\chi}{dt(t-1)}\label{bound103}\\
\geq&\dfrac{(d^2-1)n_{\gamma_i}}{t(t-1)},\label{bound104}
\end{align}
where $\rho_j$ is the reduced density matrix of $|\psi_{\gamma_i}\rangle$ on subsystem $H_j$. Equation~(\ref{bound101}) follows from linearity of the partial trace and tensor product structure. Inequality~(\ref{bound102}) follows from the expansion of the square and the positivity of cross terms. Equation~(\ref{bound103}) follows from Eq.~(\ref{bound00}) applied to subsystem $H_j$, and inequality~(\ref{bound104}) follows from the fact that $1/d \leq \mathrm{Tr}(\rho_j^2) \leq 1$. When $n_{\gamma_i}>1$, we also obtain
\begin{align}
&\sum\limits_{u=1}^{s}\sum\limits_{v=1}^{t}\Big[\textrm{Tr}\big(\mathbb{A}_{\gamma_i}^{(uv)}|\psi_{\gamma_i}\rangle\langle\psi_{\gamma_i}|\big)\Big]^2\label{}\nonumber\\
=&\sum\limits_{u=1}^{s}\sum\limits_{v=1}^{t}\Big[\sum\limits_{j\in \gamma_i}\textrm{Tr}\big(A_{j}^{(uv)}\rho_j\big)\Big]^2\label{}\nonumber\\
\leq& \sum\limits_{u=1}^{s}\sum\limits_{v=1}^{t}n_{\gamma_i}\sum\limits_{j\in
\gamma_i}\Big[\textrm{Tr}\big(A_{j}^{(uv)}\rho_j\big)\Big]^2\label{bound32}\\
\leq&\dfrac{(d-1)(d^2+t^2\chi)n_{\gamma_i}^2}{dt(t-1)}.\label{bound33}
\end{align}
Here, inequalities~(\ref{bound32}) and~(\ref{bound33}) follow from $\big(\sum_{i=1}^m x_i\big)^2\leq m\sum_{i=1}^m x_i^2$ and from Eq.~(\ref{bound000}) applied to subsystem $H_j$, respectively.
Combining the above results, for $n_{\gamma_i} > 1$, we obtain
\begin{align}
\dfrac{(d^2-1)n_{\gamma_i}}{t(t-1)}
\leq\sum\limits_{u=1}^{s}\sum\limits_{v=1}^{t}\Big[\textrm{Tr}\big(\mathbb{A}_{\gamma_i}^{(uv)}|\psi_{\gamma_i}\rangle\langle\psi_{\gamma_i}|\big)\Big]^2\label{}\nonumber
\leq\dfrac{(d-1)(d^2+t^2\chi)n_{\gamma_i}^2}{dt(t-1)}.
\end{align}
This proves inequalities~(\ref{bound10}) and~(\ref{bound3}), thus completing the proof.


\begin{thebibliography}{99}
\bibitem{HorodeckiRMP2009}
R. Horodecki, P. Horodecki, M. Horodecki, and K. Horodecki,
Quantum entanglement,
\href{https://doi.org/10.1103/RevModPhys.81.865}{Rev.\ Mod.\ Phys.\ \textbf{81}, 865 (2009).}

\bibitem{GuhneToth2009}
O. G\"uhne and G. T\'oth,
Entanglement detection,
\href{https://doi.org/10.1016/j.physrep.2009.02.004}{Phys.\ Rep.\ \textbf{474}, 1 (2009).}

\bibitem{BennettWiesnerPRL1992}
C. H. Bennett and S. J. Wiesner,
Communication via one- and two-particle operators on Einstein-Podolsky-Rosen states,
\href{https://doi.org/10.1103/PhysRevLett.69.2881}{Phys.\ Rev.\ Lett.\ \textbf{69}, 2881 (1992).}

\bibitem{BennettBrassardCrepeauJozsaPeresWoottersPRL1993}
C. H. Bennett, G. Brassard, C. Cr\'epeau, R. Jozsa, A. Peres, and W. K. Wootters,
Teleporting an unknown quantum state via dual classical and Einstein-Podolsky-Rosen channels,
\href{https://doi.org/10.1103/PhysRevLett.70.1895}{Phys.\ Rev.\ Lett.\ \textbf{70}, 1895 (1993).}

\bibitem{ShorSIAMReview1999}
P. W. Shor,
Polynomial-time algorithms for prime factorization and discrete logarithms on a quantum computer,
\href{https://doi.org/10.1137/S003614459834701}{SIAM Rev.\ \textbf{41}, 303--332 (1999).}

\bibitem{Peres1996}
A. Peres,
Separability criterion for density matrices,
\href{https://doi.org/10.1103/PhysRevLett.77.1413}{Phys.\ Rev.\ Lett.\ \textbf{77}, 1413 (1996).}

\bibitem{HorodeckiHorodeckiPRA1999}
M. Horodecki and P. Horodecki,
Reduction criterion of separability and limits for a class of distillation protocols,
\href{https://doi.org/10.1103/PhysRevA.59.4206}{Phys.\ Rev.\ A \textbf{59}, 4206 (1999).}

\bibitem{ChenWu2003}
K. Chen and L. A. Wu,
A matrix realignment method for recognizing entanglement,
\href{https://dblp.org/rec/journals/qic/ChenW03}{Quantum Inf.\ Comput.\ \textbf{3}, 193 (2003).}

\bibitem{Rudolph2003}
O. Rudolph,
Some properties of the computable cross-norm criterion for separability,
\href{https://doi.org/10.1103/PhysRevA.67.032312}{Phys.\ Rev.\ A \textbf{67}, 032312 (2003).}

\bibitem{DohertyParrilo2002PRA}
A. C. Doherty, P. A. Parrilo, and F. M. Spedalieri,
Distinguishing separable and entangled states,
\href{https://doi.org/10.1103/PhysRevLett.88.187904}{Phys.\ Rev.\ Lett.\ \textbf{88}, 187904 (2002).}

\bibitem{GuhneHyllusGittsovichEisert2007}
O. G\"uhne, P. Hyllus, O. Gittsovich, and J. Eisert,
Covariance matrices and the separability problem,
\href{https://doi.org/10.1103/PhysRevLett.99.130504}{Phys.\ Rev.\ Lett.\ \textbf{99}, 130504 (2007).}

\bibitem{GittsovichGuhneHyllusEisert2008}
O. Gittsovich, O. G\"uhne, P. Hyllus, and J. Eisert,
Unifying several separability conditions using the covariance matrix criterion,
\href{https://doi.org/10.1103/PhysRevA.78.052319}{Phys.\ Rev.\ A \textbf{78}, 052319 (2008).}

\bibitem{LiLuo2013PRA}
N. Li and S. Luo,
Entanglement detection via quantum Fisher information,
\href{https://doi.org/10.1103/PhysRevA.88.014301}{Phys.\ Rev.\ A \textbf{88}, 014301 (2013).}

\bibitem{Szalay2019Quantum}
S. Szalay,
$k$-stretchability of entanglement, and the duality of $k$-separability and $k$-producibility,
\href{https://doi.org/10.22331/q-2019-12-02-204}{Quantum \textbf{3}, 204 (2019).}

\bibitem{RenLiSmerziGessnerPRL2021}
Z. H. Ren, W. D. Li, A. Smerzi, and M. Gessner,
Metrological detection of multipartite entanglement from Young diagrams,
\href{https://doi.org/10.1103/PhysRevLett.126.080502}{Phys.\ Rev.\ Lett.\ \textbf{126}, 080502 (2021).}

\bibitem{SzalayToth2025Quantum}
S. Szalay and G. T\'oth,
Alternatives of entanglement depth and metrological entanglement criteria,
\href{https://doi.org/10.22331/q-2025-04-18-1718}{Quantum \textbf{9}, 1718 (2025).}

\bibitem{HongLuoSong2015}
Y. Hong, S. Luo, and H. T. Song,
Detecting $k$-nonseparability via quantum Fisher information,
\href{https://doi.org/10.1103/PhysRevA.91.042313}{Phys.\ Rev.\ A \textbf{91}, 042313 (2015).}

\bibitem{GaoYan2014}
T. Gao, F. L. Yan, and S. J. van Enk,
Permutationally invariant part of a density matrix and nonseparability of $N$-qubit states,
\href{https://doi.org/10.1103/PhysRevLett.112.180501}{Phys.\ Rev.\ Lett.\ \textbf{112}, 180501 (2014).}

\bibitem{EPL104.20007}
T. Gao, Y. Hong, Y. Lu, and F. L. Yan,
Efficient $k$-separability criteria for mixed multipartite quantum states,
\href{https://doi.org/10.1209/0295-5075/104/20007}{Europhys.\ Lett.\ \textbf{104}, 20007 (2013).}

\bibitem{HongGaoYanPLA2021}
Y. Hong, T. Gao, and F. L. Yan,
Detection of $k$-partite entanglement and $k$-nonseparability of multipartite quantum states,
\href{https://doi.org/10.1016/j.physleta.2021.127347}{Phys.\ Lett.\ A \textbf{401}, 127347 (2021).}

\bibitem{PlodzienChwedenczukLewensteinMieldzio2024}
M. Plodzie\'n, J. Chwede\'nczuk, M. Lewenstein, and G. Rajchel-Mieldzio\'c,
Entanglement classification and non-$k$-separability certification via Greenberger-Horne-Zeilinger-class fidelity,
\href{https://doi.org/10.1103/PhysRevA.110.032428}{Phys.\ Rev.\ A \textbf{110}, 032428 (2024).}

\bibitem{ChenPRA2005}
Z. Q. Chen,
Wigner-Yanase skew information as tests for quantum entanglement,
\href{https://doi.org/10.1103/PhysRevA.71.052302}{Phys.\ Rev.\ A \textbf{71}, 052302 (2005).}

\bibitem{Hyllus2012}
P. Hyllus, W. Laskowski, R. Krischek, C. Schwemmer, W. Wieczorek, H. Weinfurter, L. Pezz\'e, and A. Smerzi,
Fisher information and multiparticle entanglement,
\href{https://doi.org/10.1103/PhysRevA.85.022321}{Phys.\ Rev.\ A \textbf{85}, 022321 (2012).}

\bibitem{Goth2012}
G. T\'oth,
Multipartite entanglement and high-precision metrology,
\href{https://doi.org/10.1103/PhysRevA.85.022322}{Phys.\ Rev.\ A \textbf{85}, 022322 (2012).}

\bibitem{AkbariAzhdargalam2019}
Y. Akbari-Kourbolagh and M. Azhdargalam,
Entanglement criterion for multipartite systems based on quantum Fisher information,
\href{https://doi.org/10.1103/PhysRevA.99.012304}{Phys.\ Rev.\ A \textbf{99}, 012304 (2019).}

\bibitem{WuChen2025}
K. Wu, Z. H. Chen, Z. P. Xu, Z. H. Ma, and S. M. Fei,
Hybrid of gradient descent and semidefinite programming for certifying multipartite entanglement structure,
\href{https://doi.org/10.1002/qute.202400443}{Adv.\ Quantum Technol.\ \textbf{8}, 2400443 (2025).}

\bibitem{KalevGour2014}
A. Kalev and G. Gour,
Mutually unbiased measurements in finite dimensions,
\href{https://doi.org/10.1088/1367-2630/16/5/053038}{New J.\ Phys.\ \textbf{16}, 053038 (2014).}


\bibitem{Rastegin2015}
A. E. Rastegin,
On uncertainty relations and entanglement detection with mutually unbiased measurements,
\href{https://doi.org/10.1142/S1230161215500055}{Open Syst.\ Inf.\ Dyn.\ \textbf{22}, 1550005 (2015).}


\bibitem{Appleby2007}
D. M. Appleby,
Symmetric informationally complete measurements of arbitrary rank,
\href{https://doi.org/10.1134/S0030400X07090111}{Opt.\ Spectrosc.\ \textbf{103}, 416--428 (2007).}

\bibitem{GourKalev2014}
A. Kalev and G. Gour,
Construction of all general symmetric informationally complete measurements,
\href{https://doi.org/10.1088/1751-8113/47/33/335302}{J.\ Phys.\ A: Math.\ Theor.\ \textbf{47}, 335302 (2014).}

\bibitem{Rastegin2014}
A. E. Rastegin,
Notes on general SIC-POVMs,
\href{https://doi.org/10.1088/0031-8949/89/8/085101}{Phys.\ Scr.\ \textbf{89}, 085101 (2014).}

\bibitem{WiesniakPaterekZeilinger2011}
M. Wie\'sniak, T. Paterek, and A. Zeilinger,
Entanglement in mutually unbiased bases,
\href{https://doi.org/10.1088/1367-2630/13/5/053047}{New J.\ Phys.\ \textbf{13}, 053047 (2011).}

\bibitem{DurtBengtssonZyczkowski2010}
T. Durt, B. G. Englert, I. Bengtsson, and K. \.{Z}yczkowski,
On mutually unbiased bases,
\href{https://doi.org/10.1142/S0219749910006502}{Int.\ J.\ Quantum Inf.\ \textbf{8}, 535 (2010).}

\bibitem{SpenglerHuberPRA2012}
C. Spengler, M. Huber, S. Brierley, T. Adaktylos, and B. C. Hiesmayr,
Entanglement detection via mutually unbiased bases,
\href{https://doi.org/10.1103/PhysRevA.86.022311}{Phys.\ Rev.\ A \textbf{86}, 022311 (2012).}

\bibitem{ChenMaFeiPRA2014}
B. Chen, T. Ma, and S. M. Fei,
Entanglement detection using mutually unbiased measurements,
\href{https://doi.org/10.1103/PhysRevA.89.064302}{Phys.\ Rev.\ A \textbf{89}, 064302 (2014).}

\bibitem{ShenLiDuanPRA2015}
S. Q. Shen, M. Li, and X. F. Duan,
Entanglement detection via some classes of measurements,
\href{https://doi.org/10.1103/PhysRevA.91.012326}{Phys.\ Rev.\ A \textbf{91}, 012326 (2015).}

\bibitem{LiuGaoYan2018}
L. Liu, T. Gao, and F. L. Yan,
Separability criteria via some classes of measurements,
\href{https://doi.org/10.1007/s11433-017-9070-4}{Sci.\ China Phys.\ Mech.\ Astron.\ \textbf{60}, 100311 (2017).}

\bibitem{ShangAsadianPRA2018}
J. W. Shang, A. Asadian, H. J. Zhu, and O. G\"uhne,
Enhanced entanglement criterion via symmetric informationally complete measurements,
\href{https://doi.org/10.1103/PhysRevA.98.022309}{Phys.\ Rev.\ A \textbf{98}, 022309 (2018).}

\bibitem{LaiLuo2022}
L. M. Lai and S. Luo,
Detecting Einstein-Podolsky-Rosen steering via correlation matrices,
\href{https://doi.org/10.1103/PhysRevA.106.042402}{Phys.\ Rev.\ A \textbf{106}, 042402 (2022).}


\bibitem{SiudzinskaPRA2022}
K. Siudzi\'nska,
All classes of informationally complete symmetric measurements in finite dimensions,
\href{https://doi.org/10.1103/PhysRevA.105.042209}{Phys.\ Rev.\ A \textbf{105}, 042209 (2022).}

\bibitem{WangSunFeiWangQIP2024}
Z. Wang, B. Z. Sun, S. M. Fei, and Z. X. Wang,
Schmidt number criterion via general symmetric informationally complete measurements,
\href{https://doi.org/10.1007/s11128-024-04611-7}{Quantum Inf.\ Process.\ \textbf{23}, 401 (2024).}

\bibitem{TavakoliMorelliPRA2024}
A. Tavakoli and S. Morelli,
Enhanced Schmidt-number criteria based on correlation trace norms,
\href{https://doi.org/10.1103/PhysRevA.110.062417}{Phys.\ Rev.\ A \textbf{110}, 062417 (2024).}

\bibitem{LiYaoFeiFanMaPRA2024}
J. X. Li, H. M. Yao, S. M. Fei, Z. B. Fan, and H. T. Ma,
Quantum entanglement estimation via symmetric-measurement-based positive maps,
\href{https://doi.org/10.1103/PhysRevA.109.052426}{Phys.\ Rev.\ A \textbf{109}, 052426 (2024).}

\bibitem{WangFeiPRA2025}
H. F. Wang and S. M. Fei,
Symmetric measurement-induced lower bounds of concurrence,
\href{https://doi.org/10.1103/PhysRevA.111.032432}{Phys.\ Rev.\ A \textbf{111}, 032432 (2025).}

\bibitem{SiudzinskaSciRep2022}
K. Siudzi\'nska,
Indecomposability of entanglement witnesses constructed from symmetric measurements,
\href{https://doi.org/10.1038/s41598-022-14920-5}{Sci.\ Rep.\ \textbf{12}, 10785 (2022).}

\bibitem{SchumacherAlberPRA2023}
M. Schumacher and G. Alber,
Detection of typical bipartite entanglement by local generalized measurements,
\href{https://doi.org/10.1103/PhysRevA.108.042424}{Phys.\ Rev.\ A \textbf{108}, 042424 (2023).}

\bibitem{LaiLuoCommunTheorPhys.2023}
L. M. Lai and S. Luo,
Separability criteria based on a class of symmetric measurements,
\href{https://doi.org/10.1088/1572-9494/accd5b}{Commun.\ Theor.\ Phys.\ \textbf{75}, 065101 (2023).}

\bibitem{TangWuPhysScr2023}
L. Tang and F. Wu,
Enhancing some separability criteria in many-body quantum systems,
\href{https://doi.org/10.1088/1402-4896/acd151}{Phys.\ Scr.\ \textbf{98}, 065114 (2023).}

\bibitem{QiPangHouQIP2025}
X. F. Qi, Y. Y. Pang, and J. C. Hou,
Detecting genuine multipartite entanglement based on a class of symmetric measurements,
\href{https://doi.org/10.1007/s11128-025-04729-2}{Quantum Inf.\ Process.\ \textbf{24}, 117 (2025).}


\bibitem{SunLiLuo2022}
Y. Sun, N. Li and S. Luo,
Quantifying coherence relative to channels via metric-adjusted skew information,
\href{https://doi.org/10.1103/PhysRevA.106.012436}{Phys.\ Rev.\ A \textbf{106}, 012436 (2022).}

\bibitem{CaiHansen2010}
L. Cai and F. Hansen,
Metric-adjusted skew information: convexity and restricted forms of superadditivity,
\href{https://doi.org/10.1007/s11005-010-0396-2}{Lett.\ Math.\ Phys.\ \textbf{93}, 1 (2010).}

\bibitem{FrankHansen2008}
F. Hansen,
Metric adjusted skew information,
\href{https://doi.org/10.1073/pnas.0803323105}{Proc.\ Natl.\ Acad.\ Sci.\ USA \textbf{105}, 9909 (2008).}

\bibitem{HofmannTakeuchi2003}
H. F. Hofmann and S. Takeuchi,
Violation of local uncertainty relations as a signature of entanglement,
\href{https://doi.org/10.1103/PhysRevA.68.032103}{Phys.\ Rev.\ A \textbf{68}, 032103 (2003).}

\bibitem{GuhnePRL2004}
O. G\"uhne,
Characterizing entanglement via uncertainty relations,
\href{https://doi.org/10.1103/PhysRevLett.92.117903}{Phys.\ Rev.\ Lett.\ \textbf{92}, 117903 (2004).}

\bibitem{BertlmannKrammer2008}
R. A. Bertlmann and P. Krammer,
Bloch vectors for qudits,
\href{https://doi.org/10.1088/1751-8113/41/23/235303}{J.\ Phys.\ A: Math.\ Theor.\ \textbf{41}, 235303 (2008).}

\end{thebibliography}
\end{document}